\documentclass[journal,11pt,onecolumn,draftcls]{IEEEtran}
\usepackage{graphicx}
\usepackage{amssymb}

\newcommand{\vx}{\mbox{\boldmath$x$}}
\newcommand{\vy}{\mbox{\boldmath$y$}}
\newcommand{\vz}{\mbox{\boldmath$z$}}

\newcommand{\vPhi}{\mbox{\boldmath$\Phi$}}

\newcommand{\R}{\mathbb{R}}
\newcommand{\C}{\mathbb{C}}

\newcommand{\Z}{\mathbb{Z}}

\newcommand{\E}{\mbox{E}}
\newcommand{\cov}{\mbox{Cov}}

\newcommand{\var}{\mbox{Var}}
\newcommand{\cum}{\mbox{Cum}}
\newcommand{\sign}{\mbox{Sign}}

\renewcommand{\C}{{\cal C}}

\begin{document}

\title{Compressed and quantized correlation estimators}

\author{ A.~G. Zebadua, P.O. Amblard, E. Moisan  and O.J.J. Michel}


\maketitle
\begin{abstract}
In passive monitoring using sensor networks, low energy supplies drastically constrain sensors in terms of calculation and communication abilities. Designing processing algorithms at the sensor level that take into account these constraints is an important problem in this context. We study here the estimation of correlation functions between sensors using compressed acquisition and one-bit-quantization. The estimation is achieved directly using compressed samples, without considering any reconstruction of the signals. We show that if the signals of interest are far from white noise, estimation of the correlation using  $M$ compressed samples out of $N\geq M$ can be more advantageous than estimation of the correlation using $M$ consecutive samples. The analysis consists of studying the asymptotic performance of the estimators at a fixed compression rate. We provide the analysis when the compression is realized by a random projection matrix composed of independent and identically distributed entries. The framework includes widely used random projection matrices, such as Gaussian and Bernoulli matrices, and it also includes very sparse matrices. However, it does not include subsampling without replacement, for which a separate analysis is provided. 
When considering one-bit-quantization as well, the theoretical analysis is not tractable. However, empirical evidence allows the conclusion that in practical situations, compressed and quantized estimators behave sufficiently correctly to be useful in, for example, time-delay estimation and model estimation.
\end{abstract}

\IEEEpeerreviewmaketitle

 

\section{Motivation and overview}

\IEEEPARstart{T}{he} motivating application for the ideas presented in this paper was the use of  sensor networks for structural health monitoring. An example is seen in the monitoring of concrete-based structures. Sensors can be randomly embedded in concrete during the building phase of the structure, or placed on the surface of the structure. Several tasks can be performed by such a network. One such task is auto-localization, which allows the tracking of the geometry of the network, and hence the detection of changes in the geometry provoked by modifications to the medium. Another task is output-only modal identification. In these applications, estimation of correlation functions can be required ({\it e.g.}, for time-delay estimation, for power spectrum estimation).

\subsection{Constraints}

The context of this study is {\em passive} structural health monitoring. Sensors can be, for example, micro electro-mechanical system accelerometers embedded in the propagation medium or positioned on the surface of the structure. The signals of interest are elastic waves propagated in the medium that are related to uncontrolled sources, such as microseismic waves, human-activity-induced vibrations, and others \cite{VincCML14,VincCML16}.

Typical distances between neighboring sensors range from metric to decametric distances. Thus, relying on electromagnetic-wave-based active techniques for autolocalization is barely possible  \cite{PataAKHMC05}. Indeed, the relative time-delay resolution would remain very poor, and the precision (using, {\it e.g.}, received signal strength indicator-based solutions) would not discriminate enough. Electromagnetic waves  are consequently used only for transmission purposes.

In the passive framework considered here, sensors can carry out some calculations and must communicate with neighbors. However, even if some sensors in the network are highlighted as anchors ({\it i.e.}, typically wired), many of them are autonomous: their energy supply is finite ({\it i.e.}, a battery), and calculation and communication devices need to be as economical as possible.
{\em The aim of this study was to design correlation function estimators using as minimal resources as possible, in terms of both calculation and communication}. Note, however, that nowadays communication is more energy demanding than calculation and storage.

\subsection{Solutions explored}

The solutions we explore here rely on modern ideas, such as random projections, as well as old ideas, such as polarity-coincidence detectors. Indeed, we combine these ideas through the design of compressed and one-bit-quantized estimators. 

One-bit-quantized correlation estimators date back to the 1940s, with the military research for RADAR. A report published in 1966 was indeed almost entirely written during World War 2, as mentioned in its foreword \cite{VanVM66}. Polarity-coincidence estimators work on the sign of signals instead of correlating the continuous waveforms. They perform one-bit quantization prior to any estimation. In the present context of inter-sensor correlation estimation, one-bit-quantization allows the bit rate to be lowered for the transmission between sensors and to keep the energy consumption relatively low. 

As an additional means to lower data transmission requirements, we also consider compressive acquisition \cite{EldaK12}. This has been described over the last 20 years, whereby compressive sensing states that a small number of random linear combinations of a signal sample maintains the full information on the signal provided the signal is sparse in some dictionaries. More precisely, if the signal is sparse in a dictionary, it can be reconstructed exactly from compressed measurements. The reconstruction is based on optimization techniques that in general demand a lot of calculation resources. Even if there are efficient optimization procedures nowadays, such optimization techniques are too demanding in terms of resources to be considered here. 
However, it was realized in recent years that if information is present in compressed samples, it is often not necessary to reconstruct the signal if a particular task is needed, such as a classification or estimation \cite{DaveBWB10, CopaHMMD12}.

The ideas behind these developments come from the Johnson-Lindenstrauss (JL) lemma  or transform \cite{Wood14}. Depending on the context, the JL lemma or transform can be stated as the controlled approximate conservation of norms or inner products after random projection of vectors in lower dimensional spaces. As correlation is merely an inner product, we can expect that correlation can be correctly estimated after random projection. More precisely, following the details in \cite{Wood14}, a random matrix $\vPhi$ is a JL transform with parameters $\varepsilon, \delta,n$, if with probability at least $\delta$ and
  any $n$-element subset $V\subset \R^N$, $|\big<\vPhi \vx  \big| \vPhi \vy \big> -\big< \vx \big| \vy \big>|^2\leq \varepsilon \| \vx \|^2 \| \vy\|^2$ for any $(\vx,\vy)\in V^2$. When a matrix is a JL transform, it can in general be turned into a $\ell^2$-embedding of a subspace, which means that it approximately conserves the norms of all of the vectors of the subspace. For example, a matrix of size $k\times N$ with independent and identically distributed (i.i.d.) Gaussian entries 
${\cal N}(0,1/k)$ is a JL transform with parameters $\varepsilon, \delta,n$, provided that $k\geq C \varepsilon^{-2}\log(n/\delta)$ (where $C$ is a constant). Such a matrix can be shown to be a  $\ell^2$-embedding with probability $1-\delta$ for the column space of any $N\times d$ matrix $A$, provided that 
 $k\geq C \varepsilon^{-2}(d+\log(1/\delta))$ \cite{Wood14}. Then for any $\vx\in \R^d$, $(1-\varepsilon)  \| A\vx \|^2  \leq  \| \vPhi A\vx \|^2 \leq (1+\varepsilon)  \| A\vx \|^2  $.
 From this last property, it can then be shown that if $\vPhi $ is a $\ell^2$-embedding for two $N\times d$ matrices $A$ and $B$, then 
 with probability of at least $1-\delta$, $| \big<\vPhi A \vx  \big| \vPhi B \vy \big> -\big< A\vx \big| B\vy \big>|^2\leq \varepsilon \|A\vx \|^2 \| B \vy\|^2$ for any $(\vx,\vy)\in \R^d$, provided of course that $k$ satisfies the bounds given above. This (almost) preservation of the inner product is at the root of what might be called compressed (linear) processing. This was noted in \cite{DaveBWB10}, although it  appeared even earlier; {\it e.g.}, in \cite{Sarl06}.
In the following, we make use of this property, and we define and study the correlation as the inner product of compressed vectors. 
A similar problem was  studied by \cite{LiHC06} from a different perspective. This study developed here is different mainly in two points.
First, the aim of the first part of our paper is the estimation of the statistics of some vectors, and secondly, the asymptotic analysis we provide here is given at a fixed compression rate, which is not the case in \cite{LiHC06}. Finally, the study of \cite{LiHC06} was restricted to a particular class of random matrices, whereas the analysis developed here concerns any random matrices with i.i.d. entries. Interestingly, however, as in \cite{LiHC06}, we insist on the importance of the fourth-order cumulant of the entries of the random matrix. 

The approach taken here is different from correlation-matching approaches, as developed {\it e.g.} in \cite{RomeLL15} (and references therein).  In correlation matching, a correlation matrix is searched for as a linear combination of known matrices. The parameters are estimated using an optimization procedure, and the condition under which the parameters can still be estimated from compressed measurements was studied in \cite{RomeLL15}.

Coarse quantization has already been explored in the context of compressive sensing. Early references included \cite{BoufB08,GuptNR10,ZymnBC10}.
These studies then led to several developments, such as in \cite{BourAU10,JacqLBB13,PlanV13}, to cite but a few. In almost all of these studies, the problem of reconstructing a signal from one-bit-quantized compressed samples was addressed. A notable exception was \cite{GuptNR10}, in which  classification is addressed as an application. 

\subsection{Overview.}

The main results that are shown in this paper are described in the following.
The correlation between two signals $x$ and $y$ is evaluated from $N$ dimensional vectors $\vx$ and $\vy$ 
that collect successive samples of the signals. The usual correlation estimate is the inner product $c_N=N^{-1} \vx^\top \vy$, where $\top$ represents a transposition. The compressed estimator using the random matrix $\vPhi: \R^N \rightarrow \R^M,  (M\leq N)$ is defined as $C_N= (\vPhi \vx)^\top (\vPhi \vy)$. It is assumed that the entries of $\vPhi$, denoted as $\varphi_{ij}$ or $\varphi$ generically, are i.i.d.. This allows for a large choice of matrices, and even includes subsampling with replacement as a particular case. However, 
matrices $\vPhi$ used for subsampling without replacement do not satisfy the i.i.d. condition, and sampling without replacement is studied separately.

For $C_N$ to be unbiased, $\varphi$ must have zero mean and variance of $(MN)^{-1}$. 
For any $\vPhi$, we generalize for the bounds shown in \cite{AlonGMS02,AlonMS99}. These results define the bounds for the loss in variance provided by compression, and they read as
\begin{eqnarray}
\var[C_N]  -  \var[c_N] &\leq&  \frac{2}{M N^2} 
\E\big[ \| \vx \|^2_2\|\vy\|^2_2\big]  \nonumber \\&+&M   \cum_4[\varphi] \E\big[ \| \vx \circ \vy\|^2_2\big] 
\end{eqnarray}
where $\circ$ is the Hadamard product (the entry-wise product of vectors). In \cite{AlonGMS02,AlonMS99}, these bounds are given for a particular sparse matrix for which the fourth-order cumulant is negative (and is thus omitted). Interestingly, these bounds indicate that compression leads to a loss that is at most of the same order $1/M$ as the variance of the usual estimator, using $M$ consecutive samples. This holds also for very sparse matrices. We then quantify the loss (or gain) obtained by compression compared to the usual estimator calculated on $M$ consecutive samples. Indeed, compression from $N$ down to $M$ samples is interesting not only if the quality of $C_N$ is not degraded too much compared to the quality of $c_N$, but also if the quality of $C_N$ is better than the quality of the usual estimate on $M$ consecutive samples. We provide some arguments that show that the further away from white noise the signals are, the greater the advantage of compression. This means that sparsity in the spectral domain is an important hypothesis for good behavior of compressed estimates. The results are shown by studying the asymptotics $N\rightarrow +\infty$ at a fixed compression rate $\alpha=N/M$. 
The choice of the compression matrix is important. In our context of limited calculation resources, using very sparse matrices or subsampling strategies is very interesting. Compared to full matrices, such as Gaussian or Bernoulli matrices, the loss in variance is larger, although it remains reasonable. 

For the compressed and quantized estimates, the definitions are the same as before, although they are calculated using the sign of the signals.
They are given by $c^q_N=N^{-1} \sign(\vx)^\top  \sign(\vy)$ and $C^q_N=M^{-1} \sign(\vPhi \vx)^\top \sign(\vPhi \vy)$, where  the sign function applies entry wise. Even in the Gaussian case, the bias of the compressed and quantized estimator is out of reach analytically. 
This is because there is no simple closed form for the probability mass of high dimensional Gaussian vectors in an orthant. 
The problem is even more difficult for the variance. However, we argue that in many situations, some hints on the behavior of these estimators
can be given. Indeed, the compression matrix is useful, as it mixes random variables: $\vPhi \vx$  can practically be considered as Gaussian due to the central limit theorem. This is valid when the matrix is full and $\vx$ is arbitrary (the dependence structure between its components 
must be soft), or when the matrix is sparse and we restrict the signals to Gaussian. In these situations, the  mean of the quantized estimators 
is proportional to the arcsin of the correlation function targeted ({\it i.e.}, the arcsin law). 

\subsection{Organization}

The results are presented as follows. In section \ref{estimation:sec}, we first develop and study the different compressed estimators. The statistics for finite sample size are given, and then they are studied in the asymptotic regime at a fixed compression rate. The influence of the compression matrix is highlighted. As subsampling without replacement cannot be studied within the random projection framework, special treatment is devoted to it. In this section, we illustrate all of the findings by studying the AR(1) case in detail. In section \ref{quantized:sec}, we turn to the one-bit quantized version of the compressed estimates. The analysis of the quality of the estimates is essentially empirical. We conclude this section with an illustration of real data that consists of vibrations recorded in a tall building. For these measurements, we show the interest of the approach for sensor networks in structural health monitoring. All of the calculations that were developed to show the results of these studies are detailed in a separate final section. 

\section{Estimation of the correlation}
\label{estimation:sec}

\subsection{Estimators}

 Let $x(t)$ and $y(t)$ be two jointly stationary zero-mean processes. The correlation function is 
$\Gamma_{xy}(\tau)=\E[x(t) y(t+\tau)]$, and  $ C_{xyxy}(a,b,c)=\cum[x(t),y(t+a),x(t+b),y(t+c)]$  is a fourth-order cumulant based correlation function. A basic assumption is the absolute summability of these functions \cite{Bril01,BrocD91}, 
\begin{eqnarray*}
\int \big|\Gamma_{xy} (\tau) \big|\mbox{d}\tau < +\infty,\\
\int \big|C_{xyxy}(a,b,c) \big| \mbox{d}a \mbox{d}b \mbox{d}c< +\infty
\end{eqnarray*}

Two sensors labelled $x$ and $y$ deliver $N$ consecutive samples from each of the signals. The samples are stored in vectors
 $\vx=(x_{t},\ldots, x_{t-N+1})^\top$ and $\vy=(y_{t+\tau},\ldots, y_{t+\tau-N+1})^\top$, where the dependence in $t,N,\tau$ is omitted in the notation for the sake of clarity. 
 In the medium where the sensors are located, the signals $x$ and $y$ are carried by some physical waves ({\it e.g.}, acoustic, elastic). The  delay $\tau$ is the delay of propagation of the waves between sensors $x$ and $y$.
 
The usual empirical estimate of the correlation function is $c_{N,xy}(\tau)=N^{-1} \vx^\top \vy$.

To obtain a compressed estimator, the vectors $\vx$ and $\vy$ from $\R^N$  are embedded into $\R^M$, with $M\leq N$ using a random matrix. 
Let $\vPhi $ be this random matrix of dimension $M\times N$. We assume the entries 
$\varphi_{ij}$ of  $\vPhi $ are identically and independently distributed with zero mean. The distribution of the entries is not yet specified ($\varphi$ without indices
stands for a variable independent of $\varphi_{ij}$, and is distributed as $\varphi_{ij}$).
We then form the compressed estimator as $C_{N,xy}(\tau)= \left(\vPhi \vx \right)^\top \left(\vPhi \vy \right) $.
We  also consider the usual estimator evaluated on $M$ successive samples and denoted as $c_M$. Finally, we will also consider later a compressed estimator  ${\cal C}_M$
based on subsampling without replacement. This compressed estimator however does not fit into the general framework based on random embedding, and will be studied separately. 


\subsection{Statistics of the estimators}

The statistics of $c_{N,xy}(\tau)$ are well documented and can be found in any classical statistical signal-processing textbook ({\it e.g.}, \cite{Bril01,BrocD91}). To sum these up, the first-order and second-order statistics are
\begin{eqnarray}
\E[c_{N,xy}(\tau) ] &=& \Gamma_{xy}(\tau) \nonumber \\
\var [c_{N,xy}(\tau)] &=&  \frac{1}{N^2} \sum_{k=-N}^{N} \left( N-|k|\right) f_{xy}^\tau(k)
\label{StatCor:eq}
\end{eqnarray}
where $f_{xy}^\tau(k)= C_{xyxy}(\tau,k,\tau +k) + \Gamma_{xy}(\tau+k) \Gamma_{xy}(\tau-k) +\Gamma_{xx}(k)\Gamma_{yy}(k) $.
As is well known, the usual empirical estimate is unbiased and its variance has the usual $N^{-1}$ rate, provided the processes are mixing in some sense (the correlation functions rapidly decrease to zero at infinity). This condition is provided by the assumption of absolute summability made earlier, which  ensures that $f_{xy}^\tau(k)$ is summable.

The evaluation of the same statistics for the compressed estimator is not difficult, but it requires some care. 
Evaluation of the mean leads to 
$\E[C_{N,xy}(\tau)]=MN .  \E[\varphi^2] \Gamma_{xy}(\tau)$, which implies the unbiasedness condition
\begin{eqnarray}
MN.\E[\varphi^2] =1
\label{unbiasedness:eq}
\end{eqnarray}
For the variance, the calculations detailed in section \ref{proof:sec} lead to
\begin{eqnarray}
\var[C_N]  &=  &\var[c_N] +M \cum_4[\varphi] \sum_\alpha \E[x_\alpha^2 y_\alpha^2] \nonumber \\
&+&\frac{1}{M N^2} \Big(  
  \E\big[ \| \vx \|^2_2\|\vy\|^2_2\big] + \E\big[(\vx^\top \vy)^2\big]\Big)
\label{VarianceCompressedDefInter:eq}
\end{eqnarray}
where  $\cum_4[\varphi]$ is the fourth-order cumulant of $\varphi$. Some comments  can be made at this point: 
\begin{itemize}
\item The variance $\var[C_N]$  is (hopefully) greater than $\var[c_N]$;  the increment is shown to be $\E[\var[C|xy]]$ in section \ref{proof:sec}.
\item The variance depends on $\vPhi$ explicitly only through the fourth-order cumulant of its entries.
\item We can simply bound the difference of the variance:
\begin{eqnarray}
\var[C_N]  -  \var[c_N] &\leq& M \cum_4[\varphi] \sum_\alpha \E[x_\alpha^2 y_\alpha^2] \nonumber \\&+& \frac{2}{M N^2} 
\E\big[ \| \vx \|^2_2\|\vy\|^2_2\big] \big] \\
&\leq&\frac{2}{M N^2} 
\E\big[ \| \vx \|^2_2\|\vy\|^2_2\big]  
\end{eqnarray}
\end{itemize}
where the first inequality follows from the Cauchy-Schwartz inequality, and the second inequality is valid only when $\cum_4[\varphi] \leq 0$;  this is the case for matrices with Gaussian entries, uniform entries, Bernoulli entries, and several bounded random entries. 
The inequality shows in these cases that the loss incurred by compression is no more than a $1/M$ order term that is an interesting guarantee (indeed, it  is shown in section \ref{proof:sec}  that $ \frac{1}{ N^2}
\E\big[ \| \vx \|^2_2\|\vy\|^2_2\big] =O(1)$). This $1/M$ order term is of the same order as the variance of $c_M$, the usual estimator using $M$ consecutive samples. Therefore,
in the worst situations, the compressed estimator will perform as well as $c_M$, the correlation estimator evaluated on $M$ consecutive samples. We will see later, however, that it can be much more efficient than $c_M$.

To obtain the behavior of the variance as a function of $N$ and $M$, the expectations in Equation (\ref{VarianceCompressedDefInter:eq}) must be further developed. Using the developments made in section \ref{proof:sec}, we get
\begin{eqnarray}
\var[C_{N }]  &=&(1+\frac{1}{M} )   \var[c_{N }  ]\nonumber  \\
&+&MN  \cum_4 [\varphi]  \left(g_{xy}^\tau(0)+\Gamma_{xx}(0)\Gamma_{yy}(0)\right)\nonumber \\
&+&\frac{1}{M } \Big( \Gamma_{xx}(0)\Gamma_{yy}(0)+\Gamma_{xy}(\tau)^2\nonumber  \\
& & + \frac{1}{N^2} \sum_{k=-N}^{N} \left( N-|k|\right) g_{xy}^\tau(k)
 \Big) \label{VarianceCompressedDef:eq}
\end{eqnarray}
where $g_{xy}^\tau(k)= C_{xyxy}(k+\tau,0,\tau +k) +2 \Gamma_{xy}(\tau+k)^2 $.
In the following, we detail some of the consequences of these results.

\subsection{Asymptotic behavior at a fixed compression rate}

We study the estimates when $N$ and $M$ go to infinity for a fixed compression rate $\alpha=N/M\geq 1$.
The absolute summability of the second-order and fourth-order correlation function implies absolute summability of $f_{xy}^\tau(k)$ and $g_{xy}^\tau(k)$ (defined respectively in Eqs. (\ref{StatCor:eq}) and (\ref{VarianceCompressedDef:eq})). Invoking the Lebesgue dominated-convergence theorem leads to
\begin{eqnarray*}
N \var [c_{N,xy}(\tau)]&=&\sum_{k=-N}^{N} \left( 1-\frac{|k|}{N}\right) f_{xy}^\tau(k) \\
&\stackrel{N\rightarrow +\infty}{\longrightarrow}& \sum_{k\in \Z}  f_{xy}^\tau(k)  := v(\tau)
\end{eqnarray*}
Likewise, we have
\begin{eqnarray*}
N \var [c_{M,xy}(\tau)]  &=& \frac{N}{M}   \sum_{k=-M}^{M} \left( 1-\frac{|k|}{M}\right) f_{xy}^\tau(k)\\
&\stackrel{N\rightarrow +\infty}{\longrightarrow}&\alpha v(\tau)
\end{eqnarray*}
and
\begin{eqnarray*}
 \lim_{N\rightarrow +\infty} N\var[ C_{N,xy}(\tau)] &=& v(\tau)+\alpha  \left( \Gamma_{xx}(0)\Gamma_{yy}(0)+\Gamma_{xy}(\tau)^2\right) \\
 &+& \left( \Gamma_{xx}(0)\Gamma_{yy}(0)+g_{xy}^\tau(0)\right) c_{4,\varphi}
\end{eqnarray*}
where $c_{4,\varphi}= \lim_{N\rightarrow +\infty} MN^2\cum_4[\varphi] $, assuming it exists.

To compare the estimators, it is interesting to evaluate what the variance loss is between $c_{N,xy}$ and $C_{N,xy}$, and also between 
$C_{N,xy}$ and $c_{M,xy}$. Indeed, compression is interesting not only if 
\begin{eqnarray*}
\delta (C_N,c_N) &:=&\lim_{N\rightarrow +\infty} N(\var [C_{N,xy}(\tau)] -\var [c_{N,xy}(\tau)])  \\
&=& \alpha  \left( \Gamma_{xx}(0)\Gamma_{yy}(0)+\Gamma_{xy}(\tau)^2\right) \\
&+&  \left( \Gamma_{xx}(0)\Gamma_{yy}(0)+g_{xy}^\tau(0)\right) c_{4,\varphi}
\end{eqnarray*}
 is small, but also if $C_{N,xy}(\tau)$ compares favorably to $c_{M,xy}(\tau)$. Thus, we evaluate
\begin{eqnarray}
\delta (C_N,c_M)  &:=&  \lim_{N\rightarrow +\infty} N(\var [C_{N,xy}(\tau)] -\var [c_{M,xy}(\tau)] ) \nonumber \\
 &=& \delta (C_N,c_N)  \nonumber \\
 &+& \lim_{N\rightarrow +\infty} N(\var [c_{N,xy}(\tau)] -\var [c_{M,xy}(\tau)] ) \nonumber \\
&=& (1-\alpha) v(\tau) 
+ \alpha  \left( \Gamma_{xx}(0)\Gamma_{yy}(0)+\Gamma_{xy}(\tau)^2\right)\nonumber  \\
&+& \left( \Gamma_{xx}(0)\Gamma_{yy}(0)+g_{xy}^\tau(0)\right) c_{4,\varphi}
   \label{deltaC:eq}
\end{eqnarray}
Hence, if $\delta (C_N,c_M)  <0$, the compressed estimator is better than the usual estimator evaluated on $M$ points.

If the signals are jointly Gaussian and if we denote $\rho_{xy}$ as the normalized correlation function, the preceding Equation (\ref{deltaC:eq}) implies that $\delta (C_N,c_M) <0$  if and only if
\begin{eqnarray}
\sum_{k\geq 1} \Big( \rho_{xy}(\tau-k)\rho_{xy}(\tau+k)  + \rho_{xx}(k) \rho_{yy}(k)\Big)> \nonumber \\
 \frac{1+c_{4,\varphi} +\rho^2_{xy}(\tau) (1-2c_{4,\varphi})}{2(\alpha-1)}
\end{eqnarray}
In many applications, we are interested in estimating the auto-correlation function. In this case,   $y(t)=x(t)$, and the loss in variance then reads as
\begin{eqnarray*}
\frac{\delta (C_N,c_M) }{\Gamma_{xx}(0)^2}&=& (\alpha+c_{4,\varphi} )+ (\alpha +2c_{4,\varphi})\rho^2_{xx}(\tau) \\
&+&(1-\alpha)\sum_{k\in\Z} \Big( \rho_{xx}(\tau-k)\rho_{xx}(\tau+k)  + \rho_{xx}^2(k) \Big)
\end{eqnarray*}
Therefore, for a given $\tau$, the lower
$
\sum_{k\in\Z} \Big( \rho_{xx}(\tau-k)\rho_{xx}(\tau+k)  + \rho_{xx}^2(k) \Big),
$
the greater $\delta (C_N,c_M) $, and the worse the compressed estimator. 
For $\tau=0$, this expression reduces to $2\sum_{k\in\Z} \  \rho_{xx}^2(k)  $, which is always greater than or equal to  2 (because 
$\rho_{xx}(0)=1\geq |\rho_{xx}(k)| , \forall k$). Equality occurs when $\rho_{xx}(k)=\delta_k$; {\it i.e.}, when $x$ is white noise. 

When $\tau\not = 0$,   $\sum_{k\in\Z} \rho_{xx}(\tau-k)\rho_{xx}(\tau+k) $ is the convolution $\rho_{xx}\star \rho_{xx}$ evaluated at 
$2\tau$. If $\rho_{xx}$ has a very rapid decay to zero (much more rapid than $\tau$), the convolution at $2\tau$ is very small, $\delta(C_N,c_M) $ is large, and the compressed estimator  behaves poorly.

In conclusion, if the process is close to white noise, the compressed estimator behaves poorly, whereas the conclusion is reversed if the correlation 
function of the process is far from a Dirac 'function'.  We study these arguments more precisely using an AR model in subsection \ref{ARstudy:ssec}.

\subsection{Influence of the distribution of $\varphi$}

As seen in the variance expression, the distribution of  the entries of the random matrix enters through its fourth-order cumulant and is constrained to have zero mean and a second-order moment of $1/(MN)$ (unbiasedness condition of Eq. \ref{unbiasedness:eq}).  Furthermore, one of the goals in this study is to minimize local calculations as much as possible.  In light of these constraints, we discuss some different distributions.

\noindent {\bf Gaussian entries} $\varphi \sim {\cal N}(0,1/MN)$ : the advantage of this choice is to eliminate the term in  $\cum_4[\varphi]$ in the variance of the compressed estimator. A drawback when it comes to implement this choice on some chips is the 
high complexity required, both as storage capacity and calculation requirements. A Gaussian matrix is full, and obtaining the compressed vector requires 
$O(MN) $ multiplications.

\noindent {\bf Bernoulli entries} $\varphi=\pm 1/\sqrt{MN}$ equiprobably: The fourth-order cumulant is $-2(MN)^{-2}$, and therefore the  term  $MN^2 \cum_4[\varphi] $  behaves as $M^{-1}$, which leads to  $ c_{4,\varphi}=0$. Using this matrix is easy, as no multiplication is required to obtain the compressed vectors. However, the matrix is full and it requires high-capacity storage. 

\noindent {\bf Sparse matrices} : if the ternary distribution  $\varphi=\pm (2N)^{-1/2}$ with probability $M^{-1}$, and $\varphi=0$ with probability $1-2M^{-1}$, are used, $MN^2\cum_4[\varphi]=1/2-3M^{-1}$ and $c_{4,\varphi}=1/2$: this increases the variance loss.
However, the resulting matrix is very sparse, and the calculations are easy (there are a mean of $2N$ nonzero elements among the $MN$ entries of the matrix). This class of matrices was studied in a similar context in \cite{LiHC06}.

Let us note that an equivalent matrix can be used that contains exactly $N$ nonzero elements, and has the same statistical characteristics. This matrix is defined as follows.
Let $h_i$ be a stochastic process defined on $\{1,\ldots,N\}$ with values in $\{1,\ldots,M\}$. The $h_i$ are supposed to be i.i.d., and uniformly distributed. Let $\sigma_i$ be another i.i.d. stochastic process on $\{1,\ldots,N\}$, but taking values $\pm 1$ equiprobably. $\sigma$ and $h$ are assumed to be independent.  Let a matrix $\vPhi$ have entries $\varphi_{ij}= \sigma(j) \delta_{ih(j)}$: each column has only one nonzero element, chosen equiprobably in $\{1,\ldots, M\}$, the value being $\pm 1$ equiprobably.  
This matrix has zero mean entries, $\E[\varphi_{kl}]= 0$, as $\sigma$ and $h$ are independent, and $\sigma$ has zero mean. 
$\E[\varphi_{kl}^2]= 1/M$ as $\sigma$ and $h$ are independent, and $\sigma$ has variance 1, and $\E[\delta_{kh(k)}]=1/M$. Thus, to obtain the unbiasedness condition, $\vPhi/\sqrt{N}$ must be used. 
Likewise, $\E[\varphi_{kl}^4]= 1/M$ and then $\cum_4[\varphi]= (1-3/M)/M>0$. Indeed this $\vPhi/\sqrt{N}$ has the same statistics as the ternary 
$(-1/\sqrt{N},0,+1/\sqrt{N})$,
with probability law $\big(1/(2M),1-1/M,1/(2M)\big)$. The only difference is that the number of nonzero elements is $N$ almost surely. For this choice, $MN^2\cum_4[\varphi]=1-3M^{-1}$ leads to an additional term 
in the variance loss, as $ c_{4,\varphi}=1$. 

In the simulations  that  follow, the first choice of ternary distribution  $(-1/\sqrt{2N},0,+1/\sqrt{2N})$ leading to $c_{4,\varphi}=1/2$ is considered. 
However, as a practical implementation, the second choice $(-1/\sqrt{N},0,+1/\sqrt{N})$ is preferable due to its ease of implementation using tables,
and its gain of $N$ zero term (in the mean).

\noindent {\bf  Subsampling with replacement}: another simple way to compress is to randomly subsample  the vectors by sampling without or with replacement $M$ samples out of the $N$ samples of the vectors. Sampling without replacement is treated separately in section \ref{subsampwor:ssec}. For sampling with replacement, it suffices to consider the same construction made above for a very sparse matrix. 

Let $g_i$ be a stochastic process defined on $\{1,\ldots,M\}$ with values of $\{1,\ldots,N\}$. The $h_i$ are supposed to be i.i.d. and uniformly distributed. Let $\beta_i$ be another i.i.d. stochastic process on $\{1,\ldots,M\}$, but taking values $\pm 1$ equiprobably. $\beta$ and $g$ are assumed to be independent.  Let a matrix $\vPhi$ have entries $\varphi_{ij}= \beta(i) \delta_{g(i)j}$: each row has only one nonzero element (chosen equiprobably in $\{1,\ldots, N\}$), the value being $\pm 1$ equiprobably.
This matrix has zero mean entries, $\E[\varphi_{kl}]= 0$, as $\beta$ and $g$ are independent, and $\beta$ has zero mean. 
$\E[\varphi_{kl}^2]= 1/N$ as $\sigma$ and $h$ are independent, and $\sigma$ has variance 1, and $\E[\delta_{kh(k)}]=1/N$. Thus, to obtain the unbiasedness condition, $\vPhi/\sqrt{M}$ must be used. In this situation, $MN^2\cum_4[\varphi]=NM^{-1}-3M^{-1}=\alpha  -3M^{-1}$, and therefore
this choice gives an additional variance loss with $ c_{4,\varphi}=\alpha$, which can be relatively high for high compression loss (recall however that it is a $1/N$ term).

\subsection{The case of subsampling without replacement} 
\label{subsampwor:ssec}

Subsampling without replacement can be written as an embedding with a particular random matrix, although this matrix does not fulfill the hypotheses required in the framework adopted above: in a matrix $\vPhi$ under sampling without replacement, each row contains exactly one nonzero element, although no two rows can have the same nonzero element. Thus, such a random matrix cannot have independent rows, and its elements cannot be i.i.d..
A separate analysis must be made, which is detailed in section \ref{proofsSubsamp:ssec}.

The dataset is composed of $\vx=(x_1,\ldots,x_N)^\top$ and $\vy=(y_1,\ldots,y_N)^\top$. We form the Hadamard product (the entry-wise product) of the two vectors $\vz=\vx \circ \vy$.
The  subsample $Z_1,\ldots, Z_M$ is obtained uniformly at random, without replacement from $\vz$.
We form the estimator $\C_M =M^{-1}\sum_i Z_i$. We show in section  \ref{proofsSubsamp:ssec} that $\C_M$ is unbiased, and that its variance reads as
\begin{eqnarray*}
\var[\C_M] = \var[ c_N ]  + \frac{\alpha-1}{N-1} \big( f^\tau_{xy}(0)-\var[c_N] \big)
\end{eqnarray*}
where we recall that  $\alpha=N/M$ is the compression rate, and $f_{xy}^\tau(0)= C_{xyxy}(\tau,0,\tau ) + \Gamma_{xy}^2(\tau) +\Gamma_{xx}(0)\Gamma_{yy}(0) $. Therefore, asymptotically, the loss for sampling without replacement reads as
\begin{eqnarray*}
\delta (\C_M,c_N)&:=& \lim_{N\rightarrow +\infty} N(\var[\C_M] - \var[ c_N ]  ) \\&=& (\alpha-1)f^\tau_{xy}(0)
\end{eqnarray*}
For the case of sampling with replacement, we have seen that $c_{4,\varphi}=\alpha$, and we can write
\begin{eqnarray*}
\delta (C_N,c_N) = \alpha \Big( f^\tau_{xy}(0)  +2\Gamma_{xy}^2(\tau)+\Gamma_{xx} (0)\Gamma_{yy} (0)\Big)
\end{eqnarray*}
because $g_{xy}^\tau(0)=f_{xy}^\tau(0) +\Gamma_{xy}^2(\tau) -\Gamma_{xx} (0)\Gamma_{yy} (0)$. We thus see that sampling with replacement 
has an asymptotic variance loss of $ f^\tau_{xy}(0)  +\alpha \Big(2\Gamma_{xy}^2(\tau)+\Gamma_{xx} (0)\Gamma_{yy} (0)\Big)$ with respect to sampling without replacement.

\subsection{AR(1) Gaussian case}
\label{ARstudy:ssec}

The analysis developed so far shows that sparsity in the frequency domain is required to obtain good behavior of the compressed estimator. 
To illustrate this further, we consider the case of the autoregressive process of order 1. For this process, one parameter
allows the modulation of the correlation decay rate, and hence the sparsity in the frequency domain.

\begin{figure}
\begin{center}
    \includegraphics[scale=.42]{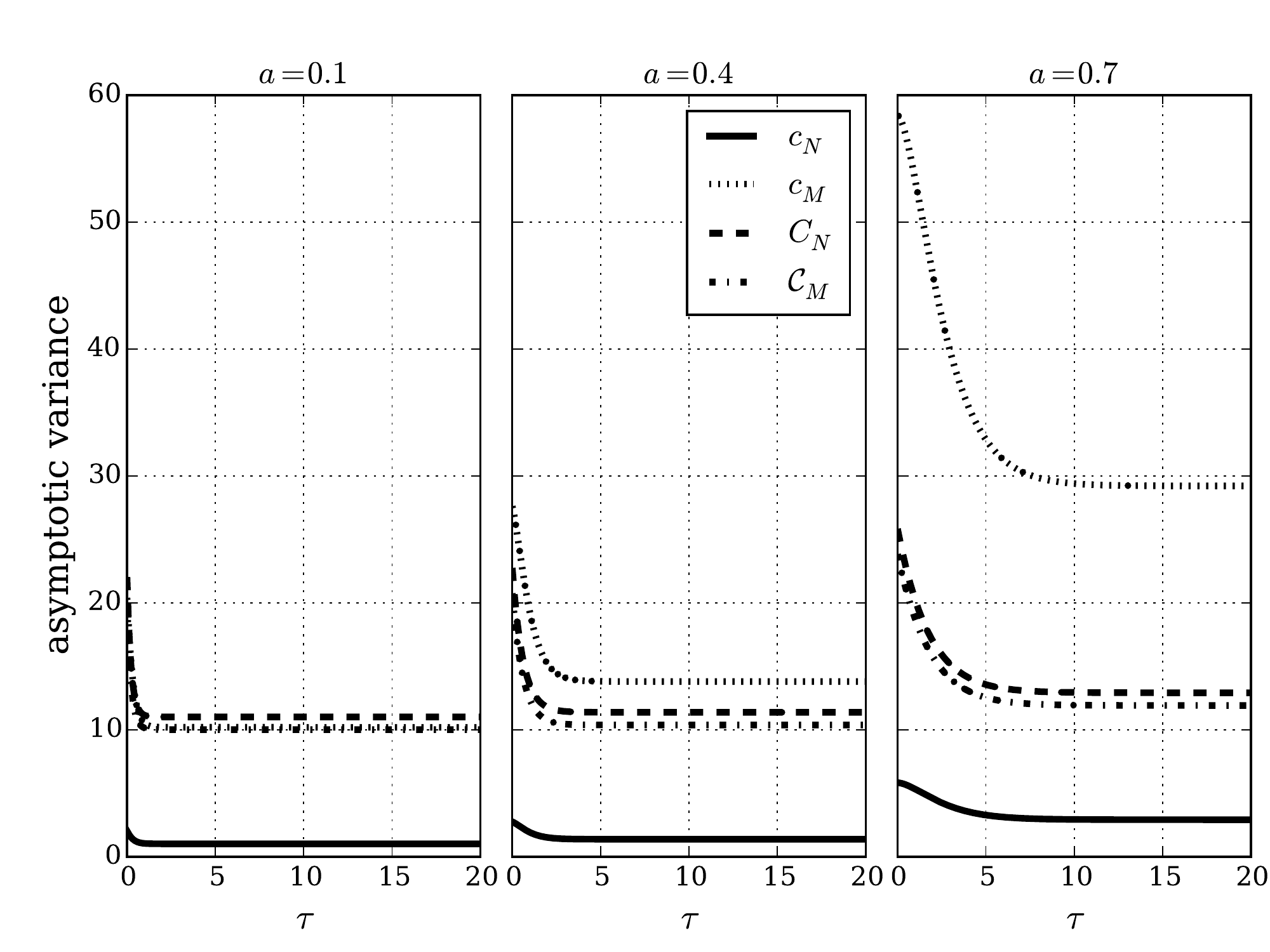}
    \end{center}
        \caption{Asymptotic variance of the estimators of the correlation function of an AR(1) process, for three different values of the leading parameter $a$, 
        for $c_N$, $c_M$, and $C_N$ with a Gaussian random matrix, and for $\C_M$, the estimator based on sampling without replacement. 
        The compressed estimator of the $N$ to $M$ samples can outperform the usual estimator based on $M$ samples if the signal is sufficiently correlated. The compression rate chosen is $\alpha=10$.}%
\label{cor_compressed_a:fig}
\end{figure}

We consider the estimation of the correlation function of  a Gaussian process that follows an AR(1) model, $x_t=a x_{t-1} + \sqrt{1-a^2} \varepsilon_t$, where $\varepsilon_t$ is a sequence of i.i.d.-normalized, zero-mean Gaussian variables. Here, $a\in (-1;1)$. The case $a=0$ corresponds to the white Gaussian noise case. We estimate  the autocorrelation function of $x$.
For this particular case, as $\Gamma_{xx}(\tau)=a^{|\tau|}$, we obtain the asymptotic variance for the different estimators
\begin{eqnarray*}
\lim_{N\rightarrow +\infty}  N \var[ c_N] &=& 1 +(2\tau+1) a^{2|\tau|} + 2a^2 \frac{1+a^{2|\tau|}}{1-a^2} \\&:=& v(\tau) \\
\lim_{N\rightarrow +\infty}  N \var[ c_M] &=& \alpha v(\tau)  \\
\lim_{N\rightarrow +\infty}  N \var[C_N] &=& v(\tau) +\alpha (1 +a^{2|\tau|} ) + c_{4,\varphi}(1 +2 a^{2|\tau|} ) \\
\lim_{N\rightarrow +\infty}  N \var[\C_M] &=& v(\tau) + (\alpha-1) (1 +a^{2|\tau|} ) 
\end{eqnarray*}
Figure (\ref{cor_compressed_a:fig}) illustrates these asymptotic variances for $a= 0, 0.4$ and 0.7, which corresponds to an increase in correlation time, and for a compression rate $\alpha=10$. The matrix $\vPhi$ that is chosen satisfies $c_{4,\varphi}=\lim_{N\rightarrow+}MN^2\cum[\varphi]=0$.  As seen in Figure (\ref{cor_compressed_a:fig}), increasing $a$ results in an improvement in $C_N$ and $\C_M$ compared to $c_M$.  As discussed earlier, the larger $|a|$   
the less white the process is, and the better the compressed estimators behave. 

\begin{figure}
\begin{center}
    \includegraphics[scale=.42]{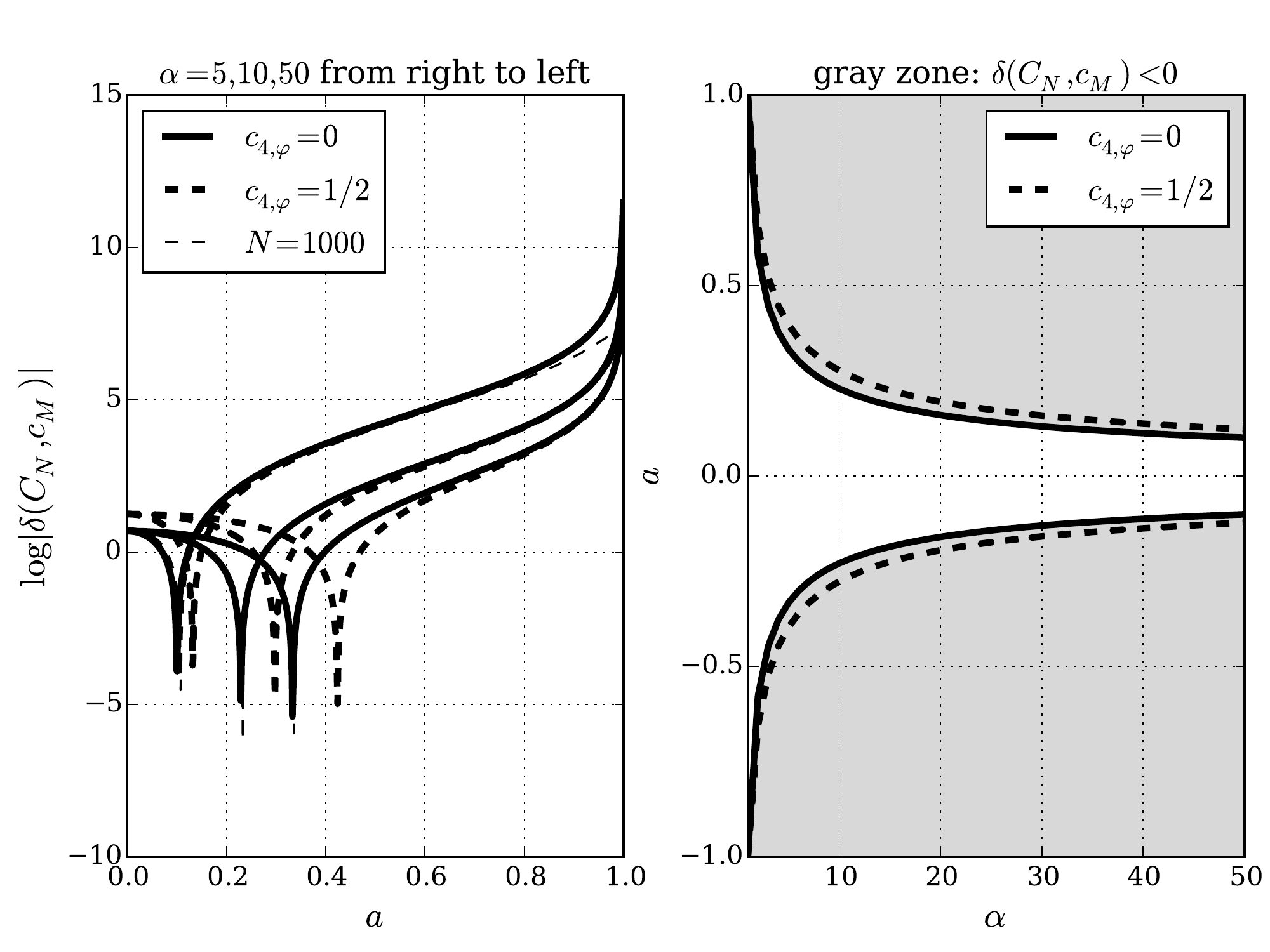} 
    \end{center}
        \caption{Left plot: $\log |\delta(C_N, c_M)| $ as a function of $a$ for three possible compression rates of $\alpha=5, 10 $ and 50 (right to left), and for $\tau=0$. 
        The thin dashed line corresponds to the evaluation of  $\log |\delta (C_N, c_M)| $  for finite $M,N$ calculations (here $N=1,000$). The thick  lines correspond  to the asymptotic limits (dashed for the ternary $\phi$, continuous for the matrices with $c_{4,\varphi}=0$ ).  For each curve, 
        the part to the left of the singularity corresponds to $\delta (C_N, c_M)>0$, whereas the right part corresponds to $\delta (C_N, c_M)<0$, for which the compressed estimator of the $N$ to $M$ samples is better than the usual estimator on $M$ samples. Right plot: Curves $|a|=f(\alpha)$ delimiting the zones of $\delta(C_N, c_M)<0 $ in gray, and the zones of $\delta (C_N, c_M)>0$ in white. The curves plotted are easily shown to satisfy $a^2> (1+3 c_{4,\varphi}/2)/( 3c_{4,\varphi}/2+2 \alpha -1)$.}%
\label{deltaM:fig}
\end{figure}

This is confirmed in Figure (\ref{deltaM:fig}), which illustrates $ \delta (C_N,c_M)$ for $\tau=0$, the difference between the variance of the compressed estimator and the usual estimator based on $M$ samples. The plot shows $   \log \big|\delta (C_N,c_M) \big|$  for three compression rates $\alpha=5, 10,$ and $50$, for two values of $c_{4,\varphi}$. Thus, the singularity in each curve corresponds to a change of sign in $\delta (C_N,c_M)$: for each curve, $\delta (C_N,c_M)>0$ is to the left of the singularity, and it is negative to the right. When $\alpha$ increases, the singularity shifts to the left. Finally, for $c_{4,\varphi}=0$ only, the nonasymptotic result is superposed, obtained here for $N=1,000$. There is good agreement between the asymptotic and nonasymptotic. Note that $\delta (C_N,c_M)$ always decreases as a function of $a$ in the asymptotic regime.  This is not the case in the nonasymptotic analysis; indeed, when $|a|=1$, the process is a constant, and random  compression cannot be better than the usual average estimation. 

The position of the singularity can be easily found in this example.
 For $\tau=0$, a rapid calculation leads to 
\begin{eqnarray*}
 \Big( \delta (C_N,c_M)  \leq 0\Big) \Longleftrightarrow   \Big(  a^2 \geq \frac{2+3c_{4,\varphi}}{4\alpha - 2 +3c_{4,\varphi}} \Big)
\end{eqnarray*}
The $(\alpha,a)$ zone where the compression is interesting is displayed in gray in the left plot of Figure (\ref{deltaM:fig}).
For   $\alpha=5, 10$ and 50,  the compressed estimator outperforms the usual estimator based on $M$ samples as soon as $a$ is large enough.  Note that in these plots the asymptotic curves are given for $c_{4,\varphi}=0$ (Gaussian or Bernoulli random matrix) as well as for $c_{4,\varphi}=1/2$ (ternary random matrix). This illustrates the 
weak influence of the matrix choice on the improvement, and opens the way to a dramatic decrease in the computational needs for a given performance. 

\subsection{Discussion.}

The effects reported here have a  simple interpretation. If the signals are white noise, the evaluation of the correlation function using $M$ consecutive samples or $M$
randomly chosen samples over a window of length $N$ will be equivalent. Furthermore, using linear combinations of these $M$ randomly chosen samples will degrade the quality of the estimation a little. In contrast, if the signals are highly correlated (in time), $M$ consecutive samples provide less information than $M$ samples chosen irregularly from $N$ consecutive samples. The interesting point used here is that this remains true if we use $M$ random linear combinations of $N$ samples.  Furthermore, as illustrated previously, the longer the correlation time the higher the gain of compression. This can be viewed as an illustration of sparsity in the frequency domain. In the example developed, the longer the correlation time, the less frequency bands occupied. It is interesting that this can also be linked to compressibility in a coding sense, as a high correlation time corresponds to a low information content and leads to a high rate coding. 

\section{One-bit-quantization-based estimates}
\label{quantized:sec}

It is well known that the statistical information content of zero crossings of a stochastic process is very close to the 
information content of the process itself. This led studies in the 1950s to implement correlation estimates
of a process using one-bit quantized measurements\footnote{This can even be traced back to the 1940s, as explained in the foreword of \cite{VanVM66}.}. This was done at that time for ease of computation using analog devices,
although this methodology has now been replaced by the usual correlation estimates due to the increase in digital computational resources. 

However,  in the era of sensor networks that demand high resources in communication, this one-bit quantization signal processing methodology
has a lot to offer. Here, we empirically demonstrate that joining these old ideas to the new ideas of compressive measurements can
dramatically decrease the need for computation and communication resources for correlation estimation and time-delay estimation in sensor networks. 

\subsection{Quantized compressed estimator}

We consider the same setting as in the previous section, except that all of the data are now one-bit quantized. For any variable $z$, we denote $\bar{z}:=\sign(z)$, the variable that is +1 if $z\geq0$ and $-1$ if $z<0$. The same notation is adopted for vectors, knowing that the operation is applied element-wise. Then, we consider the estimators ${c}^q_{N,xy}(\tau)=N^{-1} \bar{\vx}_{1}^\top \bar{\vx}_{2}$,  $C^q_{N,xy}(\tau)= \left(\overline{\vPhi \vx_{1}}\right)^\top \left(\overline{\vPhi \vx_{2}}\right) $, and $\C^q_{N,xy}(\tau)$
for the subsampling without replacement. The analysis of these estimators is more tedious than before. However, in the Gaussian case, some elements can be put forward to justify the use of these estimators. Therefore, in what follows in this section, we assume that the processes under study are jointly Gaussian.  It is well known ({\it e.g.}, \cite{Pici93,VanVM66}) that for  two normalized ({\it i.e.}, zero mean and unit variance) jointly Gaussian random variables $a$ and $b$, the correlation between their signed versions is given by 
$\E[\bar{a}\bar{b}]=(2/\pi)\arcsin(\E[ab])$. Using this result, we can evaluate the mean of the usual estimators using one-bit quantized measurements, and we get
\begin{eqnarray}
\E[c^q_{N,xy}(\tau)]& =& \frac{2}{\pi} \arcsin \frac{\Gamma_{xy}(\tau)}{\sqrt{\Gamma_{xx}(0)} \sqrt{\Gamma_{yy}(0)}} 
 \end{eqnarray}
 The mean of $\C^q_{N,xy}(\tau)$ can be easily obtained (arguments for this will be detailed shortly), and it turns out to be equal to $\E[c^q_{N,xy}(\tau)]$.
 However,  for the compressed estimator $C^q_N$,  even if $\vx$ and $\vy$ are jointly Gaussian,  the embedded vectors $\vPhi \vx$ and $\vPhi \vy$ are not jointly Gaussian, and we cannot apply the $\arcsin$ law.

 However, we can make some comments here:
\begin{itemize}
\item The compressed and quantized estimator reads $\sum_k \overline{\sum_\alpha x_\alpha \varphi_{k\alpha}} \times \overline{\sum_\alpha y_\beta \varphi_{k\alpha}}$.
When the matrix $\vPhi$ is full, which is the case for a Gaussian matrix or a Bernoulli matrix, we can expect that the variates $X_k=\sum_\alpha x_\alpha \varphi_{k\alpha}$ and $Y_k =\sum_\alpha x_\alpha \varphi_{k\alpha}$ obey jointly a central limit theorem. Indeed, under the hypothesis made, the signals $x$ and $y$ are mixing, which means that the correlation decays sufficiently fast in time, and the random matrix is independent of the signals. Thus, it is expected than when $N$ is large, we can apply the $\arcsin$ law, even if  the
signals are not Gaussian.
\item If sparse matrices are used, the preceding comment is likely to fail. In this case, when the signals are Gaussian, it  is likely that $X_k$ and $Y_k$ will remain Gaussian (conditional to the matrix). This is truly the case  with sampling without replacement. Hence in this  case again, we can apply the $\arcsin$ law. 
\item In any other case, we do not control the statistics of the quantized estimates.

\item The one-bit quantized estimators are insensitive to the power of the signals analyzed. This is reflected in the fact that they provide an estimate of the correlation function in place of the covariance function, as seen in the expression of the mean of the estimates.
\end{itemize}

Based on the previous comments, the estimators are modified to take into account the distorsion. We  define here
\begin{eqnarray*}
\widetilde{c}^q_{N,xy}(\tau) &=&\sin\Big(  \frac{\pi}{2} c^q_N\Big)\\
\widetilde{C}^q_{N,xy}(\tau) &=&\sin \Big(  \frac{\pi}{2M} C^q_M \Big) \\
\widetilde{\C}^q_M &=& \sin \Big(  \frac{\pi}{2} \C^q_M  \Big)
\end{eqnarray*}
We know from the delta method \cite{LehmC98,Lehm99} that the modified estimators behave correctly if the unmodified estimators do so; {\it i.e.}, it statisfies a usual central limit theorem. 
This is obtained if the random variables $z_i=\bar{x}_{i}\bar{y}_{i}$ form a sufficiently mixing sequence. In this case, assuming that 
$ \sqrt{N} (c^q_N - \frac{2}{\pi} \arcsin \frac{\Gamma_{xy}(\tau)}{\sqrt{\Gamma_{xx}(0)} \sqrt{\Gamma_{yy}(0)}} )$ converges in law to ${\cal N}(0,w(\tau))$, and the undistorted estimator $ \sqrt{N} ( \tilde{c}^q_{N,xy}(\tau) - \frac{\Gamma_{xy}(\tau)}{\sqrt{\Gamma_{xx}(0)} \sqrt{\Gamma_{yy}(0)}} )$
 converges to ${\cal N}\big(0,\pi^2w(\tau) (1- \frac{\Gamma_{xy}(\tau)^2}{\Gamma_{xx}(0) \Gamma_{yy}(0)} )/4\big)$. This last form is a consequence of the delta method, and
$\cos (\arcsin x)= 1-x^2$. The same result holds for $\widetilde{C}^q_N$ and $\widetilde{\C}^q_M$ if we know the variance of  $ {C}^q_N$ and ${\C}^q_M$.

Unfortunately, the variance of the one-bit quantized estimators cannot be evaluated in closed form, as there is no (known) closed form equation for the probability mass of a four-dimensional Gaussian in a positive orthant \cite{GenzB09}, except evidently in some particular cases. 
We are thus not able to give an analytic form for 
\begin{eqnarray*}
w(\tau) = \lim_{N\rightarrow +\infty} N \var[ c^q_{N,xy}(\tau) ]
\end{eqnarray*}
except in very special cases.
For more information on this particular point and its application to correlation estimation using clipping or quantization, see for example \cite{Chen68,GuptB84} and section \ref{clippingex:ssec}, where we illustrate the difficulty. We show that evaluation of the variance requires either 
numerical integration or Monte-Carlo simulation. We chose the latter.

For the compressed estimator $C^q_N$ using a random matrix, the difficulty is the same, and we cannot  evaluate the variance. For $\C^q_M$, however, it is possible to evaluate the loss due to compression with respect to $c^q_N$:  The 
 calculation of $\var[\C^q_M]$  follows the same lines as the calculation of $\var[c^q_N]$,  as if we replace $Z_i=x_i y_i$ with $Z_i=\bar{x_i} \bar{y_i}$. We show in section \ref{proofsSubsamp:ssec} that for the Gaussian case considered here, 
\begin{eqnarray*}
\E[\C^q_M] &=& \E[c^q_N] = \frac{2}{\pi} \arcsin \frac{\Gamma_{xy}(\tau)}{\sqrt{\Gamma_{xx}(0)} \sqrt{\Gamma_{yy}(0)}}\\
\var[\C^q_M] &=& \var[ c^q_N ]  +\frac{\alpha-1}{N-1} \big(  1 -\E[c^q_N]^2 -\var[c^q_N] \big)
\end{eqnarray*}
so that 
\begin{eqnarray*}
\lim_{N\rightarrow +\infty} N\big(  \var[\C^q_M] -\var[ c^q_{N,xy}(\tau) ])  =  (\alpha-1) \big(  1 - \E[c^q_N]^2  \big)
\end{eqnarray*}

\subsection{AR(1) Gaussian case.} 
\begin{figure}
\begin{center}
    \includegraphics[scale=.42]{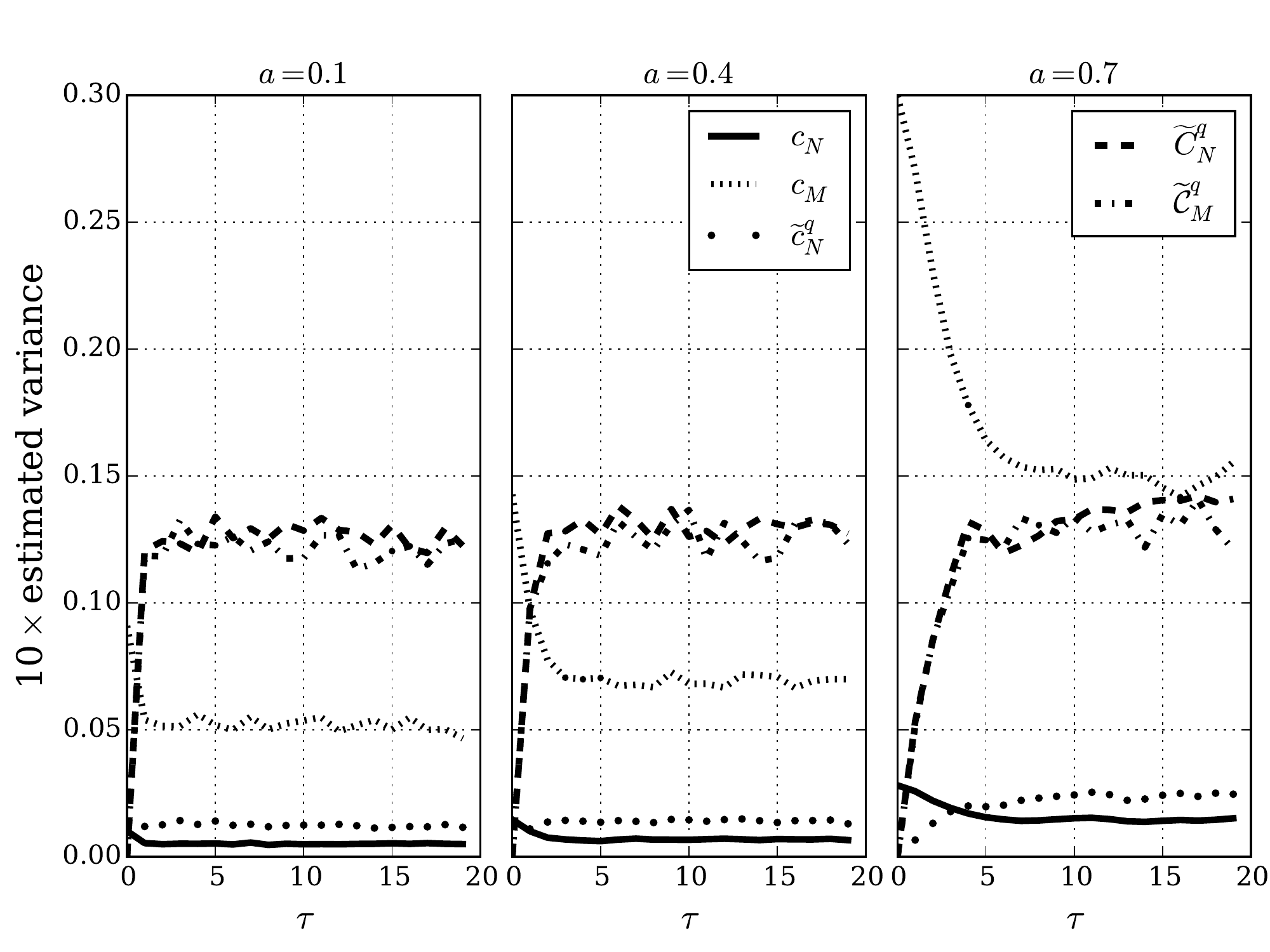}
    
    \includegraphics[scale=.42]{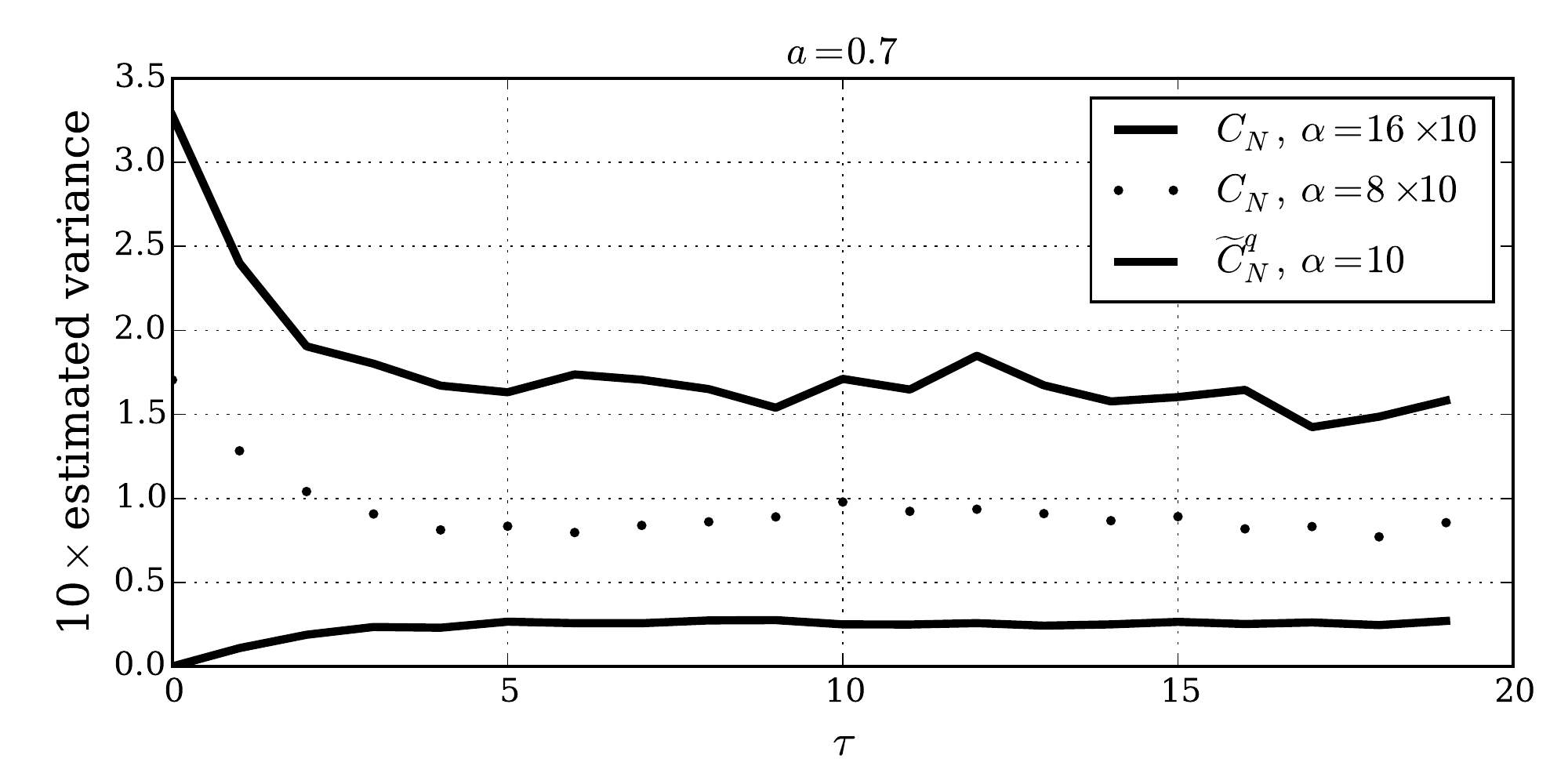}\end{center}
        \caption{Top: Variance of the estimators of the correlation function of an AR(1) process for three different values of the leading parameter.  Here, we compare the compressed estimators to their one-bit quantized versions. All of the compressed estimators were obtained using Gaussian random projection matrices. The curves were obtained by averaging over 1,000 realizations of the processes, for $N=1,000$ and $M=100$ (compression factor of 10). The one-bit quantization has a negligible effect for the usual estimator (dotted line {\it versus } continuous line). For the compressed estimators (dashed and dashed-dotted lines), quantization has more impact. However, when the process is sufficiently correlated ($a\geq0.7$, in the example), the quantized compressed estimator remains better than the usual estimator calculated over $M$ successive samples. Bottom: For the AR(1) signal with $a=0.7$, the variance of the compressed and quantized estimator for $\alpha=10$ compared to the variance of the compressed estimator for $\alpha=8\times10$ and $16\times10$. This shows that for a fixed number $M$ of transmitted bits, using the one-bit-quantized compressed estimate is preferable to using the compressed estimate. }%
\label{ARcompQuantized:fig}
\end{figure}

We apply to the AR case the same methodology ({\it i.e.}, the Gaussian random matrix) as in the preceding section, but add to the results the one-bit estimators. We illustrate the behavior of the different estimators in Figure (\ref{ARcompQuantized:fig}). For these plots, we chose $N=1,000$ and $M=100$, for which a compression factor of 10 is obtained. The correlation function of the AR process is evaluated over the first 20 lags, and the variance of the estimators is estimated by averaging over 1,000 independent snapshots of the process. 

As seen in Figure (\ref{ARcompQuantized:fig},top), we recover the elements discussed above. The larger the AR parameter, the greater the advantage of the compression. However, quantizing the signal over one bit introduces an additional distortion. For high values of the compression factor, this distortion is high, and it can double the variance ({\it e.g.}, see $a=0.4$). However, interestingly, when the process is sufficiently correlated or compressible, the loss incurred by high quantization is still compensated for by the random acquisition of $M$ samples over a horizon of $N$ samples. We note however that quantization has
a large impact at high compression rates: when comparing compressed estimators and their quantized version ({\it e.g.}, comparing 
Figs. (\ref{cor_compressed_a:fig})  and (\ref{ARcompQuantized:fig},top)   for $a=0.7$), we see that the gain obtained for the quantized version is not as large as the gain obtained using compression only.  Note, however, that by construction, quantized estimates have zero variance at the maximum of the correlation function; this is important, especially for time-delay estimation. 

The gain in variance can appear not to be that important. However, we must stress that we want to transmit as little as possible. Thus if we constraint the number of transmitted bits to M bits per correlation evaluation. If the processor used represents floats on  $f$ bits, a fair comparison would be to compare  the variance of $C_N$ for a compression rate of $ N/M= f \alpha$ to the variance of $C^q_N$ for a compression rate of $N/M=\alpha$. For  the example of AR(1) signals considered here, Figure \ref{ARcompQuantized:fig},bottom) shows these variances for $f=8$ and $f=16$, when $N$ has been set to 1,024 samples.  As seen in Figure \ref{ARcompQuantized:fig},bottom), the gain is dramatic, and for example, reaches a factor of four to eight. Therefore, for a fixed number of transmitted bits and a given required quality, using the one-bit-quantized compressed estimator is preferable to using the compressed estimator. 

\subsection{An application to real data}
In this section the methodologies developed so far are applied to real data. We take the opportunity in this application to first discuss some technological issues regarding the methods proposed.

\subsubsection{Some technological issues}
We have proposed to use random projections or subsampling to compress in the sample space, and to use one-bit quantization to compress in the amplitude space. This allows considerable gain to be obtained in terms of the calculation and transmission loads. 
For evaluation of $\Gamma_{xy}$ at sensor $x$, in the theoretical analysis we used $\vx=(x_{t-i}, i=0,\ldots, N-1)$ and
$\vy_\tau=(y_{t+\tau-i}, i=0,\ldots, N-1)$. For the compressed/ quantized estimator at sensor $x$, this $\vx$ is required, as well as  the vectors 
$\sign(\vPhi \vy_\tau)$ for all of the values of $\tau$. This requires that sensor $y$ transmits all of the vectors $\sign(\vPhi \vy_\tau)$ to sensor $x$.
This can be expensive. An alternative is to allow the sensors to have a buffer. In this case, to evaluate $\Gamma_{xy}$ at sensor $x$,
this sensor will buffer vectors $\sign(\vx_\tau)=(\sign(x_{t-\tau-i}), i=0,\ldots, N-1)$ for the values of $\tau$ required, and sensor $y$ will transmit only 
$\sign(\vPhi \vy)$, where $\vy =(y_{t-i}, i=0,\ldots, N-1)$. Thus in this set-up, if $\tau_{m}$ is the total number of lags required, the sensors
should have a buffer of at least $\tau_m M$ bits, and they have to transmit only $M$ bits to their neighbors. 
This set-up should be adopted whenever possible. Indeed, in present-day technology, the more costly part in terms of energy in sensors is the transmission. Calculation and storage capabilities are not very expensive. 

\subsubsection{Compressed and quantized estimators in action}
We consider here the real data recorded by accelerometers (SF3000L; COLIBRYS Company, www.colibrys.com) set on the ground of the 20th floor of a tall building in Grenoble, France\footnote{We thank M. Carmona and CEA/LETI in Grenoble, France, for sharing these data with us.}. For the sake of illustration, we omit any comment on the units used. Time, frequency, and amplitude are in arbitrary units.  The signal consists of 12,500 samples. 

\begin{figure}
\begin{center}
    \includegraphics[scale=.45]{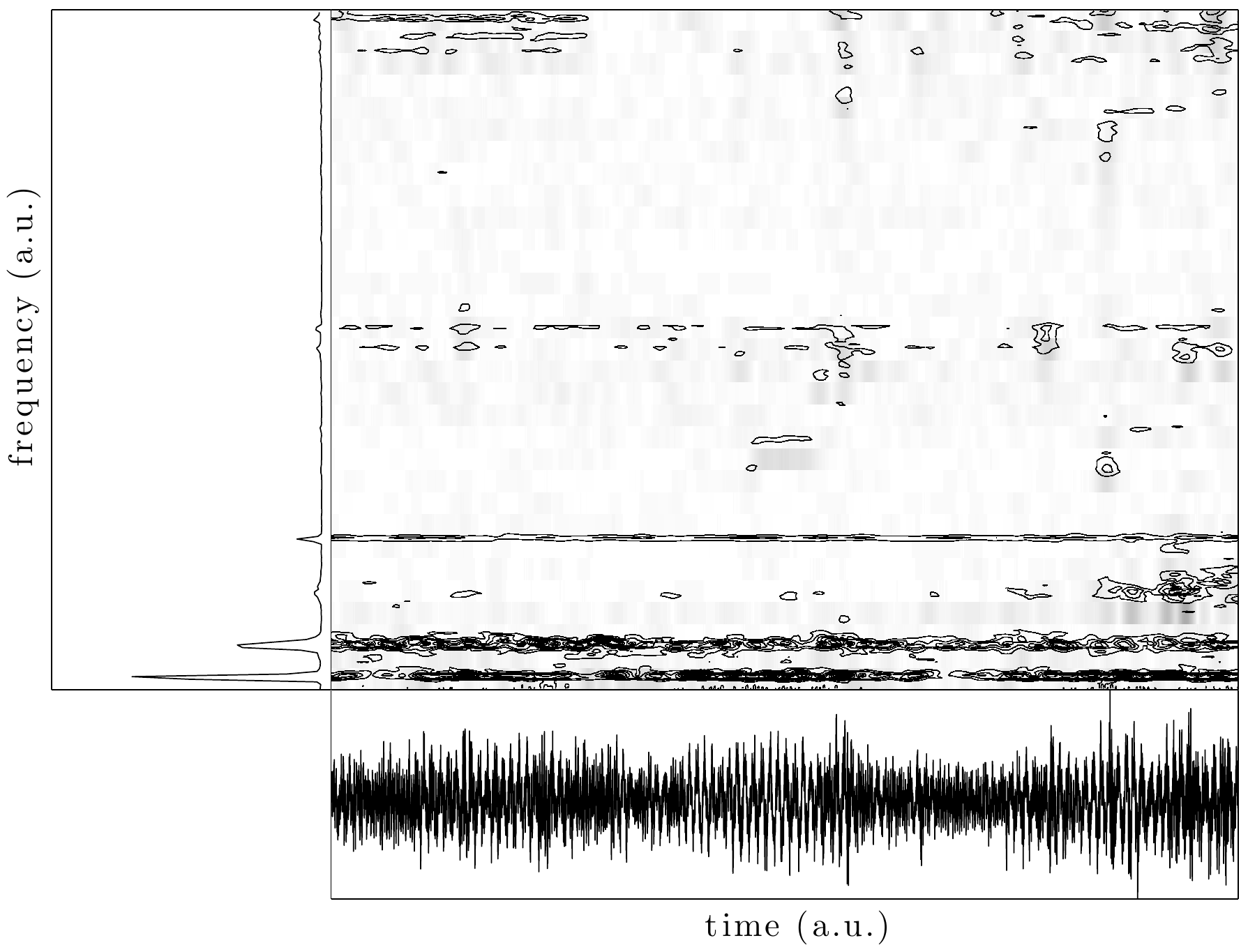}\end{center}
        \caption{Real signal used in the application. The spectrogram is shown (natural window of 500 samples, with the Fourier transform done on 512 samples; the higher the amplitude, the darker the plot; contours are added to improve contrast), with the power spectrum on the left. All of the quantities are in arbitrary units. Three well-localized harmonics show up at low frequencies. Furthermore, the spectrogram shows that stationarity is a reasonable assumption for this signal.  }%
\label{SpectroArpej:fig}
\end{figure}

The signal, its power spectrum, and a spectrogram are shown in Figure (\ref{SpectroArpej:fig}). Three well-localized harmonics show up at low frequencies. The spectrogram is illustrated to show that for the window of observation, the signal can reasonably be considered as stationary. 
Note that the signal is far from white noise, as shown by its power spectrum. Therefore, the compressed estimators are intended to behave well. 

Evaluating $\Gamma_{xx}(\tau)$ using the full dataset allows a good reference to be obtained for the correlation function. This also allows an estimate to be produced for the mean square error $\E[(\hat{\Gamma}(\tau)-\Gamma_{xx}(\tau))^2]$ for any estimator $\hat{\Gamma}(\tau)$. To study compressed and quantized estimates, the signal was cut into six blocks. The different estimators are then evaluated for each of these blocks, for $N=2,000$ samples, and $M=N/\alpha$ samples for a compression rate of $\alpha$. 

\begin{figure}
\begin{center}
    \includegraphics[scale=.5]{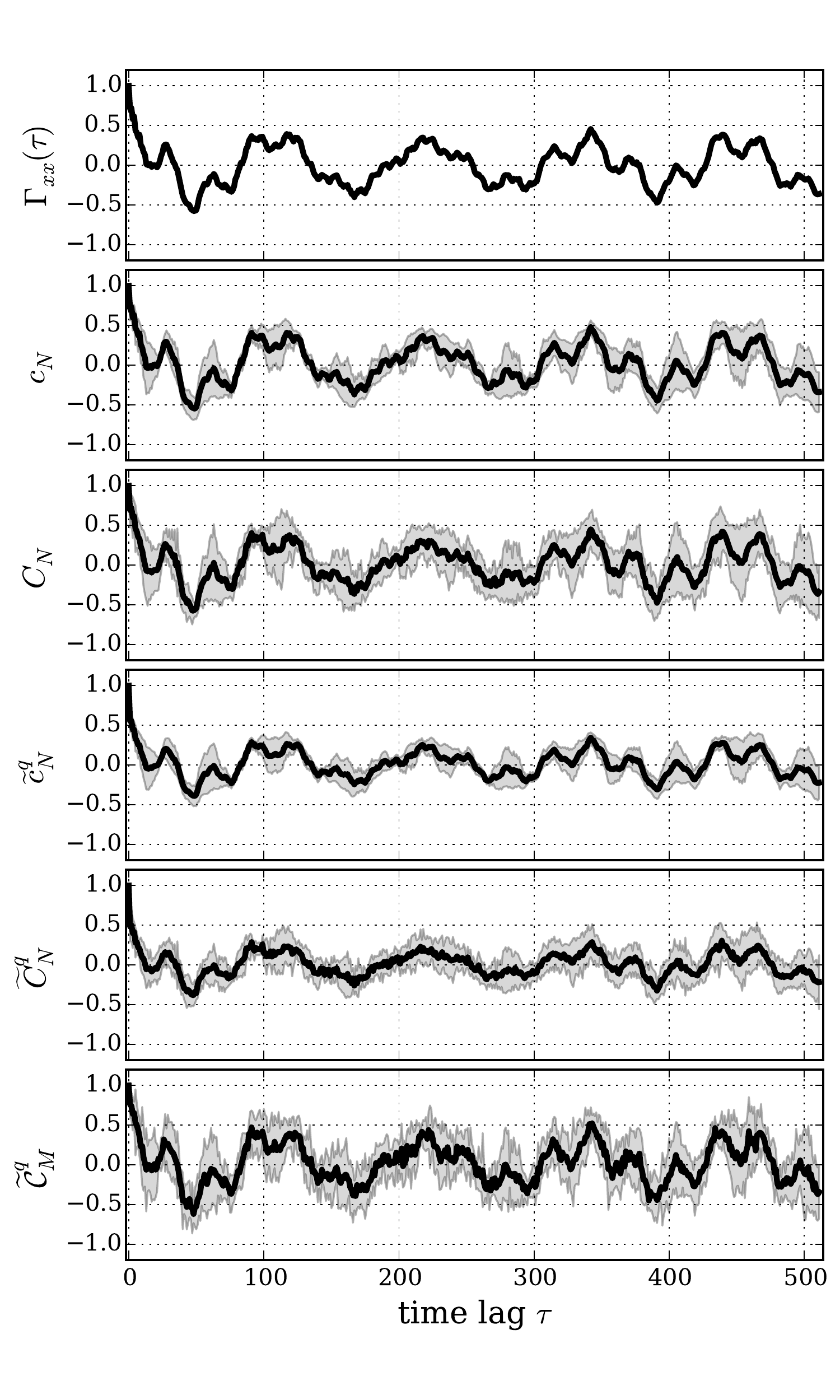}\end{center}
        \caption{Estimation of the correlation for real data. Top: The correlation function estimation for the full dataset. Second from top, to bottom: Correlation on $N$ samples $c_N$, compressed estimator $C_N$, quantized estimator $\widetilde{c}^q_N$,  compressed and quantized estimate $\widetilde{C}^q_N$ using a ternary matrix and $\alpha=10$, subsampling without replacement and quantization $\widetilde{{\cal C}}^q_M$. For these last five plots, the full dataset was cut into six blocks over which all of the estimators were applied. The functions displayed are the mean (black line) over the six blocks, $\pm$ twice the standard deviation evaluated on these six blocks (gray shading). The scales are the same for all of the plots.  }%
\label{CorArpej:fig}
\end{figure}

The different estimators are plotted in Figure (\ref{CorArpej:fig}) for $\alpha=10$, when the matrix $\vPhi$ is chosen to be very sparse ($\varphi$ is distributed according to a ternary distribution). Specifically, the estimate $\Gamma_{xx}(\tau)$ using the full dataset is depicted in the top 
plot of Figure (\ref{CorArpej:fig}), with no error bars as it is used as the ground truth. Then displayed from top to bottom there are $c_N,C_N,\widetilde{c}^q_N, \widetilde{C}^q_N$ and $\widetilde{{\cal C}}^q_M$: for each, the mean over the six blocks is plotted (continuous lines), plus/minus twice the standard deviation (gray shading around the mean) evaluated for the six blocks. This allows the mean behavior to be studied, as well as the variability over the blocks.  

Using the six blocks we also evaluate an estimate of the root mean square error (RMSE) integrated over $\tau$, taking $\Gamma_{xx}(\tau)$ as the reference.  The results are shown in Table \ref{TableEqm:tab} for  $c_N,\widetilde{c}^q_N,C_N, \widetilde{C}^q_N, \widetilde{{\cal C}}^q_M, c_M$, for 
$\alpha=5, 10,$  and 20.
 The loss in integrated RMSE for the compressed and compressed-quantized estimators is much lower than the loss of the estimators over $M$ consecutive points.  It is remarkable that the compressed-quantized estimator for $\alpha=20$ provides a good estimate of the correlation. If the performance degrades at large time lags, examination of the first 50 lags shows that in this range the estimation is very good. Furthermore, at low compression rates ($\alpha=5$), the integrated RMSE for the compressed quantized estimator is the same as that of the quantized estimate.

\begin{table}
\caption{Root mean square error (with respect to the best estimates over the whole dataset) of the estimators --quantized or not-- using $N=1,000$ samples, their compressed version for $\alpha=5, 10$ and $20$, and the usual estimator in $1,000/\alpha$ samples.}
\center \begin{tabular}{|c|c|c|c|c|c|c|}\hline  & $c_N$ & $\widetilde{c}^q_N$ &$ C_N$ & $\widetilde{C}^q_N $&$\widetilde{{\cal C}}^q_M$ & $c_M$ \\\hline\hline 
$\alpha=5$& 0.102 & 0.121 & 0.107 & 0.121 &0.157 & 0.180 \\\hline
$\alpha=10$& 0.102 & 0.121 & 0.126 & 0.147 & 0.177& 0.218 \\\hline
$\alpha=20$& 0.102 & 0.121 & 0.134 & 0.178 &0.205 & 0.238 \\\hline
 \end{tabular}
\label{TableEqm:tab}
\end{table}

\section{Concluding remarks}

In this study, old ideas are married to recent ideas on dimension reduction using random projections. We provide estimates of the correlation function between two signals by correlation of the quantized compressed acquisition of the signals. The theoretical study for the compressed part shows that compression is good when the correlation under study is far from the correlation function of white noise. In this respect, we recover the idea underlying compressed sensing, which states that  the compressed measurements  carry all of the information about a signal whenever the signal is sparse on some basis. We give a full second-order analysis of the compressed estimators. However, we only have empirical arguments for studying  the compressed and quantized estimates. The nonlinearity makes the analysis very difficult, and out of reach when thinking of closed-form equations. 
However, simulations and a real case study confirm that the  estimates proposed can be interesting within severe energy-constrained frameworks. Indeed, quantizing over one bit in amplitude, and compression at a rate of around 10 in terms of samples, leads to results that are very good when the signal has sparse spectral content. 


The theoretical study has to be pursued. The main question remains to qualify the estimators when quantization is applied. For the compressed part, we did not impose any particular model for the correlation functions. To go further in the analysis, studying the behavior of the compressed estimators for particular classes of correlations might provide more guarantees. For example, supposing that the correlation is sparse in the strict sense in the Fourier domain might be of interest. 
A second tasks is currently being developed. This consists of the evaluation of the performance of time-delay estimation based on compressed and quantized estimates. The signals that lead to good performance of the compressed estimator should be far from white noise, a property that is in contradiction with the properties required for good time-delay estimation. However, as the context here is passive monitoring, the signals used cannot be controlled, and the sources of opportunity are the only sources of information to estimate delays in the propagation.

\section{Proofs}
\label{proof:sec}

\subsection{Random matrices}
The calculations and proofs of the results given in the paper are detailed here. Recall the definitions 

$c_{N,xy}(\tau)=N^{-1} \vx^\top \vy := c $ and  $C_{N,xy}(\tau)= \left(\vPhi \vx\right)^\top \left(\vPhi \vy\right) :=C$.
Recall for later use that $x_\alpha=x(t-\alpha)$ and $y_\beta=y(t+\tau-\beta)$. Recall also that the generic entry of $\vPhi$ is denoted as $\varphi$, which can be indexed if necessary. 
We write
\begin{eqnarray*}
C &=&\sum_{k=1}^M Z_k  \mbox{ where } \\
Z_k &=& \sum_{\alpha,\beta} x_{\alpha}y_\beta \varphi_{k\alpha}\varphi_{k\beta}
\end{eqnarray*}
We first evaluate the conditional mean and variance. We have
\begin{eqnarray*}
\E[Z_k | xy] &=& \E\Big[\sum_{\alpha,\beta} x_{\alpha}y_\beta \varphi_{k\alpha}\varphi_{k\beta} \Big] \\
&=& \E[\varphi^2] \vx^\top \vy
\end{eqnarray*}
because the entries $\varphi_{ij}$ are i.i.d. and zero mean.  Thus 
$\E[C |xy]=M \E[\varphi^2] \vx^\top \vy$.

Next the conditional variance of $Z_k$ reads 
\begin{eqnarray*}
\var[ Z_k | xy] &=& \sum_{\alpha,\beta,\gamma,\delta} x_\alpha y_\beta x_\gamma y_\delta
 \cov[ \varphi_{k\alpha}\varphi_{k\beta},\varphi_{k\gamma}\varphi_{k\delta}] \\
 &=&\big(\E[\varphi^4]-\E[\varphi^2]\big)  \sum_\alpha x_\alpha^2 y_\alpha^2  \\&+& \E[\varphi^2]^2 \sum_{\alpha\not=\beta}
 \big(  x_\alpha^2 y_\beta^2 + x_\alpha y_\alpha x_\beta y_\beta \big) \\
 &=&\cum_4[\varphi] \sum_\alpha x_\alpha^2 y_\alpha^2 \\&+& \E[\varphi^2]^2 \big(   \| \vx \|^2_2\|\vy\|^2_2 + (\vx^\top \vy)^2\big)
\end{eqnarray*}
The second line is obtained because the $\varphi_{ij}$ are i.i.d. and zero mean. The sum over the four indices is then cut into four cases
$\alpha=\beta=\gamma=\delta$ and the three circular permutations of $\alpha=\beta\not=\gamma=\delta$.

Noting that $Z_k$ and $Z_l$ for $k\not=l$ are independent conditionally to $x,y$,  we obtain
\begin{eqnarray*}
\var[ C|xy] &=&M \cum_4[\varphi] \sum_\alpha x_\alpha^2 y_\alpha^2 \\&+&M \E[\varphi^2]^2 \big(   \| \vx \|^2_2\|\vy\|^2_2 + (\vx^\top \vy)^2\big)
\end{eqnarray*}

Finally, we get 
\begin{eqnarray}
\E[C]&=&E_{xy}\big[\E[C|xy]\big]  \nonumber \\
&=&M \E[\varphi^2] \E[\vx^\top \vy]  \label{meanofC:eq}\\
\var[C] &=&   E_{xy}\left[ \var[ C | xy ] \right] + \var_{xy}\left[ \E[C  | xy ] \right]  \nonumber \\
&=&M^2\E[\varphi^2]^2 \var [\vx^\top \vy] +M \cum_4[\varphi] \sum_\alpha \E[x_\alpha^2 y_\alpha^2] \nonumber
\\&+&M \E[\varphi^2]^2 \Big(  
\E\big[ \| \vx \|^2_2\|\vy\|^2_2\big] + \E\big[(\vx^\top \vy)^2\big]\Big) \label{varianceofC:eq}
\end{eqnarray}

The empirical estimate based on $N$ samples is given by $c=\vx^\top\vy/N$ and is unbiased. Thus, for $C$ to be unbiased, 
examining Equation (\ref{meanofC:eq}) shows that the variance of $\varphi$ must satisfy $\E[\varphi^2]=1/(MN)$. Imposing this unbiasedness condition in Equation (\ref{varianceofC:eq}) leads to 
\begin{eqnarray}
\var[C]  &= & \var[c] +M \cum_4[\varphi] \sum_\alpha \E[x_\alpha^2 y_\alpha^2] \nonumber\\&+&\frac{1}{M N^2} \Big(  
\E\big[ \| \vx \|^2_2\|\vy\|^2_2\big] + \E\big[(\vx^\top \vy)^2\big]\Big)
\label{VarianceInterm:eq}
\end{eqnarray}

To obtain Equation (\ref{VarianceCompressedDef:eq}), recall that $x_\alpha=x(t-\alpha)$ and $y_\beta=y(t+\tau-\beta)$, define  $ C_{xyxy}(a,b,c)=\cum[x(t),y(t+a),x(t+b),y(t+c)]$ 
and $g_{xy}^\tau(k)= C_{xyxy}(k+\tau,0,\tau +k) +2 \Gamma_{xy}(\tau+k)^2 $ to evaluate
\begin{eqnarray*}
 \sum_\alpha \E[x_\alpha^2 y_\alpha^2]  &=& N \big( C_{xyxy}(\tau,0,\tau) + 2\Gamma_{xy}(\tau)^2+ \Gamma_{xx}(0)\Gamma_{yy}(0) \big) \\
 &=& N \big( g_{xy}^\tau(0)+ \Gamma_{xx}(0)\Gamma_{yy}(0)\big) \\
 \E\big[ \| \vx \|^2_2\|\vy\|^2_2\big] &=& \sum_{\alpha,\beta} \E[  x_\alpha^2 y_\beta^2 ]  \\
 &=& \sum_{\alpha,\beta}  \big( C_{xyxy}(\tau+\alpha-\beta,0,\tau+\alpha-\beta) \\&+& 2\Gamma_{xy}(\tau+\alpha-\beta)^2+ \Gamma_{xx}(0)\Gamma_{yy}(0)\big)\\
 &=&N^2 \Gamma_{xx}(0)\Gamma_{yy}(0) +  \sum_{k=-N}^N (N-|k|) g_{xy}^\tau(k)  \\
\E\big[(\vx^\top \vy)^2\big]&=&\var \big[(\vx^\top \vy)\big]
+\E\big[(\vx^\top \vy)\big]^2 \\
  &=&N^2 \var \big[c\big]+ N^2 \Gamma_{xy}(\tau)^2
\end{eqnarray*}
where the last expression holds since the empirical estimator is an unbiased estimate of the correlation function.
Plugging the last three expressions in Equation (\ref{VarianceInterm:eq}) leads to Equation (\ref{VarianceCompressedDef:eq}).

\subsection{Subsampling without replacement}
\label{proofsSubsamp:ssec}

Let $\vx=x_1,\ldots,x_N$ and $\vy=y_1,\ldots,y_N$ be the data. Let $\vz=\vx \circ \vy$ be the Hadamard product (entry-wise product) of the two vectors.

We sample uniformly at random without replacement $M$ elements from $\{1,\ldots,N\}$. Successive samples are obtained independently. Let $S_M$ be the subsample obtained. Then, $i \in S_M$ with probability $\pi_1=M/N$, and for $i\not=j, (i,j)\in S_M$ with probability $\pi_{2}=M(M-1)/(N(N-1))$.

The estimator can be written as $\C_M=M^{-1} \sum_{i\in S_M} z_i$ or equivalently $\C_M=M^{-1} \sum_{i=1}^N z_i \varepsilon_i$ where
$\varepsilon_i, i=1,\ldots,N$ is series of Bernoulli variables of   parameter $\pi_1$. These variables are correlated, and their correlation is given by  
$E[\varepsilon_i \varepsilon_j] = \pi_{2}$. They are supposed to be independent from the $z_i$. 

Recall that  $c_N=N^{-1}\sum_{i=1}^N z_i $ is unbiased. Then, $ E[\C_M]=M^{-1}\sum_{i=1}^N E[z_i E[\varepsilon_i|z_i]] = E[c_N]$ which shows that $C_M$ is an unbiased estimator of the correlation. 

The calculation of the variance of $\C_M$ makes use of the law of total covariance, written for any random elements $X,Y,Z$ as $\cov[X,Y]=E_Z[\cov[X,Y|Z]]+\cov[E[X|Z],E[Y|Z]]$ . We have
\begin{eqnarray*}
\var[\C_M]&=&\frac{1}{M^2} \sum_{i,j} \cov[z_i\varepsilon_i,z_j\varepsilon_j] \\
&=& \frac{1}{M^2} \sum_{i,j}  E\big[z_iz_j \cov[\varepsilon_i,\varepsilon_i]\big]  + \frac{1}{M^2} \sum_{i,j}  \pi_1^2\cov[ z_i,z_j] 
\end{eqnarray*}
where the law of total covariance has been applied, and where we have used the independence between the $\varepsilon_i$s and the $z_i$s.
Since $\pi_1=M/N$, the second sum in the last expression is equal to $\var[c_N]$. Then, cutting the first sum into two parts we get
\begin{eqnarray*}
\var[\C_M]&=&\var[c_N] + \frac{1}{M^2}\sum_i \pi_1(1-\pi_1) E[z_i^2] \\&+& \frac{1}{M^2}\sum_{i\not=j} (\pi_{2} -\pi_1^2) E[z_iz_j] 
\end{eqnarray*}
Replacing $\pi_1$ by $M/N$ and $\pi_{2}$ by $M(M-1)/(N(N-1))$, using the rate of compression $\alpha=N/M$ then leads to 
\begin{eqnarray*}
\var[\C_M]&=&\var[c_N] + \frac{\alpha-1}{N-1}\Big( \frac{N-1}{N^2}\sum_i  E[z_i^2] - \frac{1}{N^2}\sum_{i\not=j}   E[z_iz_j]   \Big)\\
&=&\var[c_N] + \frac{\alpha-1}{N-1}\Big( \frac{1}{N}\sum_i  E[z_i^2] - \frac{1}{N^2}\sum_{i,j}   E[z_iz_j]   \Big)\\
&=& \var[c_N] + \frac{\alpha-1}{N-1}\Big( \var[x_1y_1] - \var[c_N]  \Big)\\
\end{eqnarray*}
where stationarity of the $z_i$s has been used. 
If the sequence $z_i$ is i.i.d., then 
$
\var[\C_M] = \var[ c_N ]  + \frac{\alpha-1}{N-1} \frac{N-1}{N} \var [z ]
$
and we recover the simple expression  $\var [\C_M]= \alpha \var[c_N] $. Indeed, selecting $M$ samples out of $N$ i.i.d. samples
leads to this result immediately. Furthermore, back to the estimation of $\Gamma_{xy}(\tau)$, we have
\begin{eqnarray*}
\var[\C_M] = \var[ c_N ]  + \frac{\alpha-1}{N-1} \big( f^\tau_{xy}(0) -\var[c_N] \big)
\end{eqnarray*}
where we can recall that $f_{xy}^\tau(k)= C_{xyxy}(\tau,k,\tau +k) + \Gamma_{xy}(\tau+k) \Gamma_{xy}(\tau-k) +\Gamma_{xx}(k)\Gamma_{yy}(k) $.

Note that nowhere do we use the distribution of $x$ and $y$. Therefore, the calculation remains valid if we work on the quantized signals.
Let $\bar{x}=\sign(x)$ , $\bar{y}=\sign(y)$, $\vz=\bar{\vx} \circ \bar{\vy}$,   $c^q_N = N^{-1}\bar{\vx} \bar{\vy}$ and $\C^q_M= M^{-1} \sum_{i\in S_M} z_i$, where the sample
$S_M$ is taken uniformly at random without replacement from $\{1,\ldots,N\}$. Then, from the results above, we have
\begin{eqnarray*}
\E[\C^q_M] &=& \E[c^q_N] = \frac{2}{\pi} \arcsin \frac{\Gamma_{xy}(\tau)}{\sqrt{\Gamma_{xx}(0)} \sqrt{\Gamma_{yy}(0)}}\\
\var[\C^q_M] &=& \var[ c^q_N ]  +\frac{\alpha-1}{N-1} \big( \var[ \bar{x_1} \bar{y_1}] -\var[c^q_N] \big)
\end{eqnarray*}
However, we can easily evaluate $\var[ \bar{x_1} \bar{y_1}] $ as 
\begin{eqnarray*}
\var[ \bar{x_1} \bar{y_1}]  &=& \E[\bar{x_1} \bar{y_1}\bar{x_1} \bar{y_1}]-\E[\bar{x_1} \bar{y_1}]^2\\
&=& 1 - \frac{4}{\pi^2} \arcsin^2 \frac{\Gamma_{xy}(\tau)}{\sqrt{\Gamma_{xx}(0)} \sqrt{\Gamma_{yy}(0)}}
\end{eqnarray*}
the last line of which is valid under the Gaussian assumption.

\subsection{Variance of the quantized estimator when $x=y$}
 \label{clippingex:ssec}

In \cite{Chen68,GuptB84}, the variance of $c^q$ is detailed in a particular case.
The existence of closed form solutions that might be of interest here are obtained only in simple cases.  To illustrate this, in the particular case $x(t)=y(t)$ for which we denote $\rho(\tau)=\rho_{xx}(\tau)$ as the normalized correlation function, the results obtained in \cite{GuptB84}  lead to 
\begin{eqnarray*}
\var[c^q_N]&=&\frac{1}{N} \sum_{k=-N}^N (1-\frac{|k|}{N} ) ( 2I^\tau_1(k)+I^\tau_2(k)+I^\tau_3(k) ) \\ & -& \E[c^q_{N,xy}(\tau)]^2
\end{eqnarray*}
where the $I^\tau_i(k)$ are defined as follows. Let $\lambda_{ij}$ be the entries of a four-dimensional correlation matrix $\Lambda$ (normalized), and 
let $c_{ij}$ be the entries of the partial correlation matrix (normalized)  ({\it i.e. } $-\Lambda_d^{-1/2}\Lambda^{-1}\Lambda_d^{-1/2}$, where 
$\Lambda_d$ is the diagonal matrix extracted from $\Lambda^{-1}$). When $x=y$, the matrix $\Lambda$ is given by 
\begin{eqnarray}
\Lambda=\left(\begin{array}{cccc}
1 & \rho(\tau) & \rho(k) & \rho(\tau+k) \\
  & 1 & \rho(\tau-k)&  \rho(k)  \\
   &   & 1 &\rho(\tau)\\
     &   &   & 1\end{array}\right)
     \label{matrixLambda:eq}
\end{eqnarray}
Then the three terms $I^\tau_i(k)$ read 
\begin{eqnarray*}
I^\tau_1(k)&= &\int_0^{\rho(\tau)}  \frac{\arcsin (c_{34}) d\lambda_{12}}{\sqrt{1-\lambda_{12}^2}}\\
I^\tau_2(k)&= &\int_{\rho(k)}^{\rho(\tau+k)}  \frac{\arcsin (c_{23}) d\lambda_{14}}{\sqrt{1-\lambda_{14}^2}}
\end{eqnarray*}
and  $I^\tau_3(k)= I^\tau_2(-k)$. In these equations, the coefficient $c_{34}$ (resp. $c_{23}$) is a nonlinear function of the $\lambda_{ij}$ defined in Equation (\ref{matrixLambda:eq}), except obviously for $\lambda_{12}$ (resp. $\lambda_{14}$), which is the dummy variable of integration. 
We thus see that even for $x=y$ the evaluation of the variance requires numerical integration or Monte-Carlo simulation. We chose the latter.

\bibliographystyle{unsrt}

\end{document}